\documentclass[twocolumn,showpacs,fleqn,nobibnotes]{revtex4}

\usepackage{amsmath}
\usepackage{graphicx}
\usepackage{float}
\usepackage{subfigure}

\def\lsim{\raise0.3ex\hbox{$<$\kern-0.75em\raise-1.1ex\hbox{$\sim$}}}
\def\gsim{\raise0.3ex\hbox{$>$\kern-0.75em\raise-1.1ex\hbox{$\sim$}}}

\begin{document}

\title{Diffractive dissociation in proton-nucleus collisions at collider energies}
\author{G. Sampaio dos Santos and M.V.T. Machado}

\affiliation{High Energy Physics Phenomenology Group, GFPAE  IF-UFRGS \\
Caixa Postal 15051, CEP 91501-970, Porto Alegre, RS, Brazil}

\begin{abstract}

The cross section for the nuclear diffractive dissociation in proton-lead collisions at the LHC is estimated. Based on the current theoretical uncertainties for the single (target) diffactive cross section in hadron-hadron reactions one obtains  $\sigma^{pA{\rightarrow}pX}_{SD} (\sqrt{s}=5.02 \,\mathrm{TeV})=19.67\pm 5.41$ mb  and  $\sigma^{pA{\rightarrow}pX}_{SD} (\sqrt{s}=8.8 \,\mathrm{TeV})=18.76\pm 5.77$ mb, respectively.
The invariant mass $M_{X}$ for the reaction $pPb \rightarrow pX$ is also analyzed. Discussion is performed on the main theoretical uncertainties associated to the calculations. 

\end{abstract}

\maketitle

\section{Introduction} 

The advent of the Large Hadron Collider (LHC) opened a new window for the studies on diffraction, elastic and inelastic scattering as they are not strongly contaminated by non-diffractive events. This is translated in the Regge theory language saying that the  scattering amplitude is completely determined by a Pomeron exchange. The current measurements on the single and double diffraction at the LHC in proton-proton collisions are in very good shape, covering the energies of 0.9 TeV \cite{ALICE}, 2.76 TeV \cite{ALICE} and  7 TeV \cite{TOTEM,CMS,ALICE}. As a general aspect, they are compatible with several available theoretical models. On the other hand, dissociation processes in which nuclei are the diffractively excited objects are far to be easily understood. The experimental constraints in these cases are given by low energy fixed-target experiments \cite{HELIOS,ehsna22,HERAB} and it was found that measured target-diffraction cross sections have quite distinct atomic mass dependences when different beam particles are considered ($h=p,\,\pi^+,\,K^+$). In general, in those experiments only the single diffractive dissociation of the target nucleus (TSD), $pA\rightarrow pX$, is measured. From theoretical point of view, the TSD cross section at high energies is a fraction of the quasi-elastic scattering cross section, i.e. the process $pA\rightarrow pA^*$ where the nucleus is excited.  The theoretical understanding is still unclear as the dual parton model \cite{Ranft} overestimates the HELIOS  data. A more optimistic approach is given by Ref. \cite{Batista} where the diffractive dissociation in nuclei  in proton-nucleus and meson-nucleus collisions is analyzed and it is able to explain the different atomic mass dependences as measured by HELIOS \cite{HELIOS} and EHS/NA22 \cite{ehsna22} with the effect of using distinct beam hadrons. Finally, in Refs. \cite{KPI,ACKPI} the cross sections for a variety of diffractive processes in proton-nucleus ($pA$)  are computed within the Glauber-Gribov theory, where inelastic shadowing corrections are summed to all orders by using the dipole representation. Their predictions overestimate the HERA-B data around 30 \% in the case of TSD cross section. Thus, an analysis on the TSD cross section in the LHC regime is quite timely.

In the present work, we investigate the single diffractive dissociation of the target nucleus (TSD) at the LHC energies in proton-lead collisions. We think this is quite relevant for the diffraction physics groups at LHC collaborations. The paper is organized as follows. In next section we summarize the theoretical information to compute the differential cross section $d^2\sigma /dtdM_X$ as a function of center of mass energy, $\sqrt{s}$, and atomic number, $A$.  In section \ref{discuss} we present the numerical calculations and discuss the main theoretical uncertainties and a comparison to other approaches is done. We cross check the consistency of phenomenological approach against the low energy fixed target experiments on proton-nucleus collisions. Finally,  we show the main conclusions.

\section{Theoretical framework and phenomenological applications}

Let us summarize the main expressions related to the theoretical approach to be used in the current analysis. The invariant cross section of a process of the type  $h p \rightarrow h^{\prime} X$  for 
hadron-proton collisions is written as
\begin{equation}
E\frac{d^3\sigma}{d{\bf p}^3} =
\frac{s}{\pi} \frac{d^2\sigma}{dt\ dM^{2}_{X}},
\label{eq:1}
\end{equation}
where $M_{X}$ is the invariant mass of diffractively excited system. Often, in literature one uses the variable $\xi = M^{2}_{X}/s$ to describe such process. In the Regge theory, in the single pole approximation, the single-diffractive cross section for producing a high-mass system, $M_X$, is dominated by the triple-Regge diagrams. At very high energies, the process is dominated by the ($I\!\!PI\!\!P)I\!\!P$ and ($I\!\!PI\!\!P)I\!\!R$ terms. The later, dominates the process at very low mass and vanishes at higher masses. The single diffractive cross section, Eq.~(\ref{eq:1}), is then expressed in terms of the triple Regge limit \cite{collins} and we take into account the dominant one, ($I\!\!PI\!\!P)I\!\!P$. We notice that the rich resonance structure in the low mass region is ignored in present work (for details on the complete treatment of low $M_X$ region see, e.g. Refs. \cite{Laszlo1,Laszlo2}). Here, the approach is supplemented by a phenomenological procedure in order to take into account corrections that guarantee unitarity \cite{goulianos}:
\begin{equation}
\frac{d^{2}\sigma_{\mathrm{SD}}}{d\xi\ dt} (h p \rightarrow h X) = 
f_{R}^h (\xi,t) \times \sigma_{{\tt I\!P} p}(M_X^2=s\,\xi) 
\label{eq:2}
\end{equation}
where the function $f_{R}^h$ is the {\it renormalized} Pomeron flux factor. A comment is in order here: it is well known that the unitarity bound can be easily broken in Regge models with the Pomeron intercept above unity, and that unitarity, or absorption corrections, restore  it. These correction have been studied in numerous works for the total and diffractive cross section. We have chosen the ad hoc parametrisation of Ref. \cite{goulianos} for its simplicity and easy numerical implementation.

Following Ref. \cite{goulianos}, this flux is constructed based on the the {\it standard} flux factor given by the Donnachie-Landshoff 
expression \cite{donnachie}, $f_{S}^h (\xi,t)$. Namely, 
\begin{eqnarray}
f_{R}^h(\xi,t)& = & \frac{1}{N(s)} \frac{\beta^{2}_{h}}{16\pi}\ F^{2}(t)\ 
\xi^{[1-2\alpha_{{\tt I\!P}}(t)]},
\label{eq:3} \\
N(s)& = & \int_{\xi_{min}}^{\xi_{max}}\int_{-\infty}^{0}f_{S}^h (\xi,t)\ dt\ d\xi , 
\label{eq:4}
\end{eqnarray}
where $\xi_{min}=1.5/s$ and $\xi_{max}=0.1$. The {\it standard} flux factor is obtained from Eq.~(\ref{eq:3}) putting $N(s)=1$. Here, $\beta_{p}=6.82$ $GeV^{-1}$ is the hadron-pomeron coupling at the quasi-elastic vertices (see Ref. \cite{covolan}). The quantity $F(t)$ is the hadron form factor, in general modeled for proton case by the Dirac form factor.
%
The Pomeron-proton cross section, appearing in  Eq.~(\ref{eq:1}), is written
as  
\begin{equation}
\sigma_{{\tt I\!P}p}(M^{2}_{X})=\beta_{p}\ g_{{\tt I\!P}}\ (s\ \xi)^{\epsilon},
\label{eq:6}
\end{equation}
with $g_{{\tt I\!P}}=0.87$ $GeV^{-1}$ and the Pomeron trajectory $\alpha(t)=1+\epsilon+\alpha't$. In our numerical calculation we take $\epsilon=0.104$ and $\alpha'=0.25$ $GeV^{-2}$ as obtained in Ref. \cite{covolan}. Finally, the diffractive cross section for the process $p p \rightarrow p X$ is obtained by integrating Eq. (\ref{eq:1}) in the experimental limits for $t$ and $M_X$ variables.

Concerning diffraction in nuclear collisions, the invariant cross section for the reaction $h A \rightarrow h X$ is expressed as follows \cite{frichter}:
\begin{equation}
\frac{d^{3}\sigma}{dx\ dp^{2}_{T}}(hA{\rightarrow}hX) = 
\sum_{m=1}^{A}\ \sigma^{hA}_{m}\ D^{N}_{m}(x,p^{2}_{T}),
\label{eq:7}
\end{equation}
where the distribution $D^{N}_{m}(x,p^{2}_{T})$ is generated by a recurrence 
formula using the assumption \cite{batista1}:
\begin{eqnarray}
D_{m=1}^{N}(x,p^{2}_{T})=
\frac{1}{\sigma^{hp}_{inel}}\ 
\left(\frac{d^3\sigma}{dx\ dp^{2}_{T}}\right )^{hN\rightarrow pX}.
\label{dsigpp}
\end{eqnarray}

The quantity $\sigma^{hA}_{m}$ is the partial inelastic cross section 
resulting of $m$ interactions, which is written in terms of the probability, $P_{A}(b)$, of a nucleon to suffer an inelastic 
interaction at a given impact parameter $b$: 
\begin{equation}
\sigma^{hA}_{m}=\int d^{2}b\ \frac{A!}{m!(A-m)!}\ P^{m}_{A}(b)\ 
[1-P_{A}(b)]^{A-m},
\label{eq:8}
\end{equation}
 where $P_{A}(b)$ is expressed in terms of the nuclear density, $\rho_{A}(z,b)$, 
\begin{equation}
P_{A}(b)=\sigma_{inel}^{hp}\int_{-\infty}^{+\infty} dz\ \rho_{A}(z,b).
\label{prob}
\end{equation}

Concerning the nuclear densities, for light nuclei ($A \leq 12$) we have used a Gaussian distribution with the width properly adjusted to the radius. Following Ref. \cite{Rybicki}, we consider $\rho_{A}(\vec r) =\rho_{max}\exp (-r^2/2R_0^2)$, with $R_0=R/\sqrt{2\ln 2}$ and the half-density radius adjusted to $R=1.8885$ fm. For heavier nuclei ($A\geq 12$), $\rho(r)$ is calculated according 
to the Woods-Saxon formula \cite{barret}. For the inelastic cross sections $\sigma_{inel}^{hp}$ in Eq.~(\ref{prob}) we have considered the parameterization presented in Ref. \cite{BH}, which describes properly the accelerator data.

When only diffractive events are taken into account, it is enough to put 
$m=1$ as these processes are supposed to take place through 
single peripheral interactions with the outlying nucleons \cite{Batista}. The last factor in square brackets in Eq. (\ref{eq:8}) controls the impact parameter dependence and makes the process peripheral e naturally cuts the large $m$ contribution. Therefore, based on this argument one uses Eqs.~(\ref{eq:7}) and (\ref{dsigpp}) to write down the single diffractive component of the invariant 
cross section in hadron-nucleus collisions as 
\begin{equation}
\frac{d^{2}\sigma_{\mathrm{SD}}}{d\xi\ dt}(h A{\rightarrow}h X) = S^2_{pA}(s,A)
\,\frac{d^{2}\sigma_{\mathrm{SD}}}{d\xi\ dt}(hp{\rightarrow}hX),
\label{dsdmnuc}
\end{equation}  
 with $S^2_{pA}(s,A)=\sigma^{hA}_{m=1}/\sigma^{hp}_{inel}$ being a function weakly dependent on energy. Integrating Eq.~(\ref{dsdmnuc}) over $\xi$ and $t$, one obtains for the single diffractive hadron-nucleus cross section:
\begin{equation}
\sigma_{\mathrm{SD}} (hA{\rightarrow}hX) = S^2_{pA}(s,A) \,\sigma_{\mathrm{SD}}(hp{\rightarrow}hX).
\label{Eq.16}
\end{equation} 

In next section we will perform the numerical calculation using expressions above in hadron-hadron and  proton-nucleus collisions, including the comparison to low energy fixed target measurements.

\section{Results and Discussion}
\label{discuss}


Before presenting the predictions for the single diffractive cross section at $pA$ collisions at the LHC, we will check the reliability of the current model. The description of low energy and Tevatron data was already tested in Ref. \cite{goulianos}.
As a cross-check, we consider the  ALICE data \cite{ALICE} at $\sqrt{s}=0.9, \,2.76$ and 7 TeV ($M_X<200$ GeV/c$^2$) and preliminary data from TOTEM \cite{TOTEM} ($3.4<M_X<1100$ GeV/c$^2$) and CMS \cite{CMS} ($12<M_X<394$ GeV/c$^2$) as well. The model fairly describes data within the error bars and it is consistent with the analysis presented by the ALICE Collaboration \cite{ALICE}. Imposing the same mass-cut as used in LHC experiments we obtain $\sigma_{\mathrm{SD}}^{\mathrm{theo}}(M_X<200) = 6.67$ mb, $\sigma_{\mathrm{SD}}^{\mathrm{theo}}(3.4<M_X<1100) = 6.48$ mb and $\sigma_{\mathrm{SD}}^{\mathrm{theo}}(12<M_X<394) = 3.79$ mb at $\sqrt{s}=7$ TeV. These results can be compared to the measured values $12.2\pm 6.6$ mb (ALICE), $6.5\pm 1.3$ mb (TOTEM) and $4.27\pm 0.87$ mb (CMS), respectively.

\begin{figure}[t]
\includegraphics[scale=0.47]{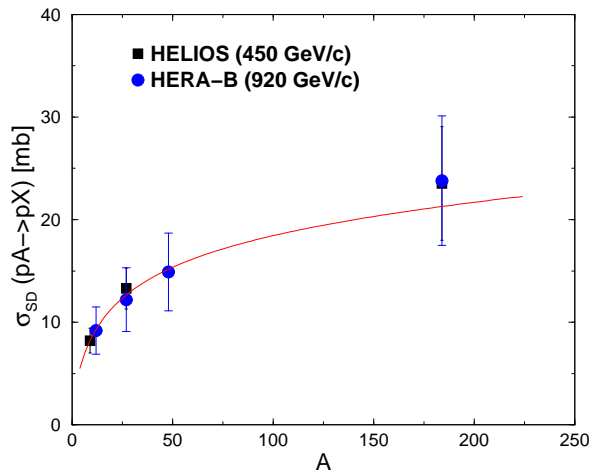}
\caption{(Color online) The data for single-diffractive cross section for the reaction $pA\rightarrow pX$ for several nuclear targets from HELIOS \cite{HELIOS} and HERA-B \cite{HERAB} collaborations. The theoretical curve (solid line) is obtained from Eq. (\ref{dsdmnuc}) taking the experimental cuts from HELIOS analysis.}
\label{fig:2}
\end{figure}

Now, the second step is to test the reliability in describing the proton-nucleus data. In Fig. \ref{fig:2} is shown the prediction for the diffractive dissociation as a function of atomic mass number. We have compared it to the low energy proton-nucleus data available from HELIOS \cite{HELIOS} (Be, Al, W) and HERA-B \cite{HERAB} (C, Al, Ti, W) experiments. The curve is obtained using the HELIOS experimental cuts $0.01\leq |t|\leq 0.36$ (GeV/c)$^2$ and $(M_X^2/s) \leq 0.075$. The model provides a good description for the $A$ dependence in proton-nucleus case. A comparison  of the same approach using $\pi^+$ and $K^+$ beams instead of protons has been done in Ref. \cite{Batista} and it nicely describes the EHS/NA22  data \cite{ehsna22}. Thus, the phenomenological model provides a fairly good description for soft nuclear diffraction. In the referred low-energy experiments only the single diffractive dissociation of the target nucleus (TSD), $pA\rightarrow pX$, is measured. Theoretically, the TSD cross section at high energies is a fraction of the quasi-elastic scattering cross section, where the nucleus is excited. Along these lines, in Refs. \cite{KPI,ACKPI} the TSD cross section has been computed using Glauber-Gribov theory including inelastic corrections to all orders of the multiple interaction. It was shown in \cite{KPI} that the simple Glauber model overestimates the HERA-B data in around 30 $\%$. The inclusion of inelastic correction introduces a relative sensibility to the model-dependence and it is able to describe reasonably the data. For sake of comparison, we contrast our results and those from Ref. \cite{ACKPI} for carbon and wolfran. We obtain $\sigma_{\mathrm{TSD}}^{\mathrm{theo}}(pA\rightarrow pX )= 9.15$ mb and 20.7 mb whereas in Ref. \cite{ACKPI}  it was found $\sigma_{\mathrm{TSD}}^{\mathrm{theo}}(pA\rightarrow pX)= 12.7 \pm 0.2$ mb and $34.7 \pm 1.0$ mb, respectively. The measured values by HERA-B are $\sigma_{\mathrm{TSD}}^{\mathrm{exp}}(pC\rightarrow pX)= 9.2 \pm 2.3$ mb and $\sigma_{\mathrm{TSD}}^{\mathrm{exp}}(pW\rightarrow pX)= 23.8 \pm 6.3$ mb.

\begin{figure}[t]
\includegraphics[scale=0.45]{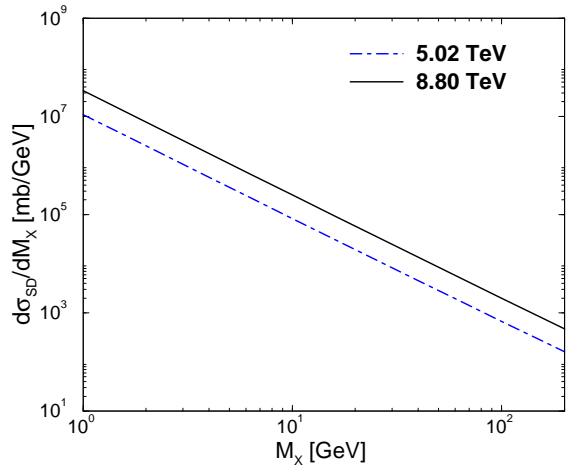}
\caption{(Color online) Differential cross section $d\sigma_{\mathrm{SD}}/dM_X$ for the reaction $pA\rightarrow pX$, integrated over $t$, for the nuclear target lead. The dot-dashed line corresponds to the energy of $\sqrt{s_{pA}}=5.02$ TeV and the solid line stands for $\sqrt{s_{pA}}=8.8$ TeV, respectively.}
\label{fig:3}
\end{figure}

Finally, let us perform predictions for the cross section in proton-lead collisions at the LHC. It is considered the previous centre of mass energy of 5.02 TeV and also 8.8 TeV (the design $pA$ energy). In Fig. \ref{fig:3} we show the differential cross section as a function of the invariant mass $M_{X}$ for the reaction $pPb \rightarrow p+ X$ using the cut $|t|\leq 1$ (GeV/c)$^2$. Using the conservative cut $M_X^2\leq 0.05 s$ the theoretical predictions for the cross section  $\sigma^{pA{\rightarrow}pX}_{\mathrm{SD}}$ are 13.9 mb and 15.4 mb for $\sqrt{s} = 5.02 $ TeV and 8.8 TeV, respectively. On the other hand, the theoretical uncertainty can be estimated using the ALICE analysis on $pp$ collisions (see Fig. 11 of Ref. \cite{ALICE}), where a comparison among distinct theoretical models was performed. Using an extrapolation from that analysis, we obtain  $\sigma^{pA{\rightarrow}pX}_{\mathrm{SD}} (\sqrt{s}=5.02 \,\mathrm{TeV})=19.67\pm 5.41$ mb  and  $\sigma^{pA{\rightarrow}pX}_{\mathrm{SD}} (\sqrt{s}=8.8 \,\mathrm{TeV})=18.76\pm 5.77$ mb. Therefore, the results presented here can be then considered a lower bound for the single diffractive production.  The present calculation can be directly compared to predictions in Ref. \cite{ACKPI}, where it was found $\sigma_{\mathrm{TSD}}^{\mathrm{theo}}(pPb\rightarrow pX )= 17.59 \pm 0.04$ mb. The agreement is good and in Ref, \cite{ACKPI} the TSD cross section is obtained from the cross section for  quasi-elastic process, $pA\rightarrow pA^*$, in the following way:
\begin{eqnarray}
\sigma_{\mathrm{TSD}}(pA\rightarrow pX )= \frac{\sigma_{\mathrm{sd}}^{pp}}{\sigma_{\mathrm{el}}^{pp}}\sigma_{\mathrm{QEL}}(pA\rightarrow pA^*)
\end{eqnarray}

The expression above allow us to calculate predictions for the quasi-elastic cross section. Using the parameterization for the elastic cross section from Ref. \cite{BH}, we obtain $\sigma_{\mathrm{QEL}}(pA\rightarrow pA^*)=S^2_{pA}(s,A)\,\sigma_{\mathrm{el}}^{pp}= 
 80.0$ mb and  80.5 mb (for 5.05 TeV and 8.8 TeV, respectively). This is slightly lower than 112.65 mb obtained in Ref. \cite{ACKPI} using Glauber approach and nucleon-nucleon short range correlations.

As a last analysis, we need to estimate the single diffractive beam component (BSD), $pA\rightarrow XA$. In  Ref. \cite{FMS} the inelastic coherent nuclear diffraction is investigated and the role played by color fluctuations in the projectile wave function was demonstrated. The expression for the diffractive cross section for BSD is non-trivial in general case but an approximated expression which is qualitatively good for all values of $A$ is given by \cite{FMS}:
\begin{eqnarray}
\sigma_{\mathrm{BSD}}(pA\rightarrow XA)\simeq \frac{\omega_{\sigma} \,\langle \sigma  \rangle^2}{4}\int d^2b\,T^2_A(b)\,e^{-\langle \sigma \rangle \,T_A(b)},\nonumber \\
\end{eqnarray}
where $\omega_{\sigma}  \equiv [\langle \sigma^2 \rangle-\langle \sigma \rangle^2  ]/\langle \sigma \rangle^2 $  (with $\langle \sigma \rangle = \sigma_{tot}^{pp}$) and $T_A$ is the nuclear profile function. Following Ref. \cite{GSPLB}, we consider $\omega_{\sigma}=0.1$ for the LHC energies and the total cross section from \cite{BH} (we checked that the approximated expression is a factor 2 below the full calculation from \cite{GSPLB} for lead nuclei).  We found the following values: $\sigma_{\mathrm{BSD}}(pA\rightarrow XA) = 11.8$ mb and 11.5 mb for $\sqrt{s} = 5.02$ and 8.8 TeV, respectively. This is order of magnitude compared to the TSD cross section computed here.

In summary, we have presented a phenomenological analysis on the diffractive cross section of proton-nucleus collisions.  Based on the current theoretical uncertainties on the diffractive hadron-hadron cross section, we obtain the following values for the single (target) dissociation cross sections,  $\sigma_{\mathrm{TSD}}(pA\rightarrow pX) =19.67\pm 5.41$ mb  and  $18.76\pm 5.77$ mb at $\sqrt{s}=5.02$ TeV and 8.8 TeV. For the quasi-elastic cross section  we obtain $\sigma_{\mathrm{QEL}}(pA\rightarrow pA^*)\simeq 80$ mb in the same energy range, whereas for the single (beam) dissociation cross section it was found  $\sigma_{\mathrm{BSD}}(pA\rightarrow XA) \simeq 12$ mb.

\section*{Acknowledgments}
This work was  partially financed by the Brazilian funding
agency CNPq and by the French-Brazilian scientific cooperation project CAPES-COFECUB 744/12. MVTM thanks the kind hospitality at IPhT CEA Saclay (France), where this work has started.

\end{document}